\documentclass{article}
\usepackage[utf8]{inputenc}
\usepackage[english]{babel}

\usepackage[numbers,super]{natbib}
\usepackage[pdftex]{graphicx}
\usepackage{caption}
\usepackage{amsmath,amssymb}
\usepackage[]{pdfpages}
\usepackage{enumerate}
\usepackage[normalem]{ulem}
\usepackage{hyperref}

\renewcommand{\deg}{\ensuremath{^\circ}}
\def\deg{\ifmmode{^\circ}\else{$^\circ$}\fi}
\def\h2o{\ifmmode{{\rm H}_2{\rm O}}\else{H$_2$O}\fi}

\usepackage{lineno}
\def\imagetop#1{\vtop{\null\hbox{#1}}}

\begin{linenomath*}
\end{linenomath*}

\begin{document}

\title{Micro cold traps on the Moon}
\date{}

\maketitle



\author{P. O. Hayne\footnote{Laboratory for Atmospheric \& Space Physics, and Astrophysical \& Planetary Sciences Department, University of Colorado Boulder, Colorado, USA. \href{mailto:Paul.Hayne@Colorado.edu}{Paul.Hayne@Colorado.edu}}}
\author{O. Aharonson\footnote{Helen Kimmel Center for Planetary Science, Weizmann Institute of Science, Rehovot, Israel.}$^{,3}$}
\author{N. Sch\"{o}rghofer\footnote{Planetary Science Institute, Tucson, Arizona, USA.}$^{,}$\footnote{Planetary Science Institute, Honolulu, Hawaii, USA}}




\begin{abstract}
Water ice is thought to be trapped in large permanently shadowed regions (PSRs) in the Moon's polar regions, due to their extremely low temperatures. Here, we show that many unmapped cold traps exist on small spatial scales, substantially augmenting the areas where ice may accumulate.
Using theoretical models and data from the Lunar Reconnaissance Orbiter, we estimate the contribution of shadows on scales from 1~km down to 1~cm, the smallest distance over which we find cold-trapping to be effective for water ice. Approximately 10--20\% of the permanent cold trap area for water is found to be contained in these ``micro cold traps," which are the most numerous cold traps on the Moon. Consideration of all spatial scales therefore substantially increases the number of cold traps over previous estimates, for a total area of $\sim$40,000~km$^2$.
A majority of cold traps for water ice is found at latitudes $>80$\deg\ because permanent shadows equatorward of 80\deg\ are typically too warm to support ice accumulation.
Our results show that water trapped at the lunar poles may be more accessible as a resource for future missions than previously thought.
\end{abstract}

\section{Introduction}
Water is unstable on much of the lunar surface, due to the high temperatures and rapid photo-destruction under direct solar illumination. However, water ice and other volatiles are thought to be trapped near the Moon's poles, where large permanently shadowed regions (PSRs) exist due to the lunar topography and the small spin axis obliquity to the Sun \citep{urey52, WMB61a}. In some of the polar PSRs, temperatures are low enough \citep{vasavada99} ($<$110~K) that the thermal lifetime of ice may be longer than the age of the solar system; these are termed `cold traps'. Water delivered to the lunar surface may eventually become cold-trapped at the poles in the form of ice. A similar process is thought to operate on Mercury and Ceres, where large ice deposits have been found \citep{harmon92,harmon11,platz17} in  the locations predicted by thermal models \citep{vasavada99,neumann13,paige13,ermakov17}. So far, evidence for similar ice deposits on the Moon has been inconsistent \citep{margot99,feldman02,colaprete10}, despite a strong theoretical basis for their existence \citep{WMB61b,arnold79}, and concerted efforts to locate and quantify them \citep{chin2007lunar}. The highly inhomogeneous distribution of lunar resources may also result in difficulties in implementing the Outer Space Treaty that declared the Moon as providence of all humankind \citep{elvis16}.

\begin{figure*} [th]
\centering
\imagetop{\includegraphics[width=\textwidth]{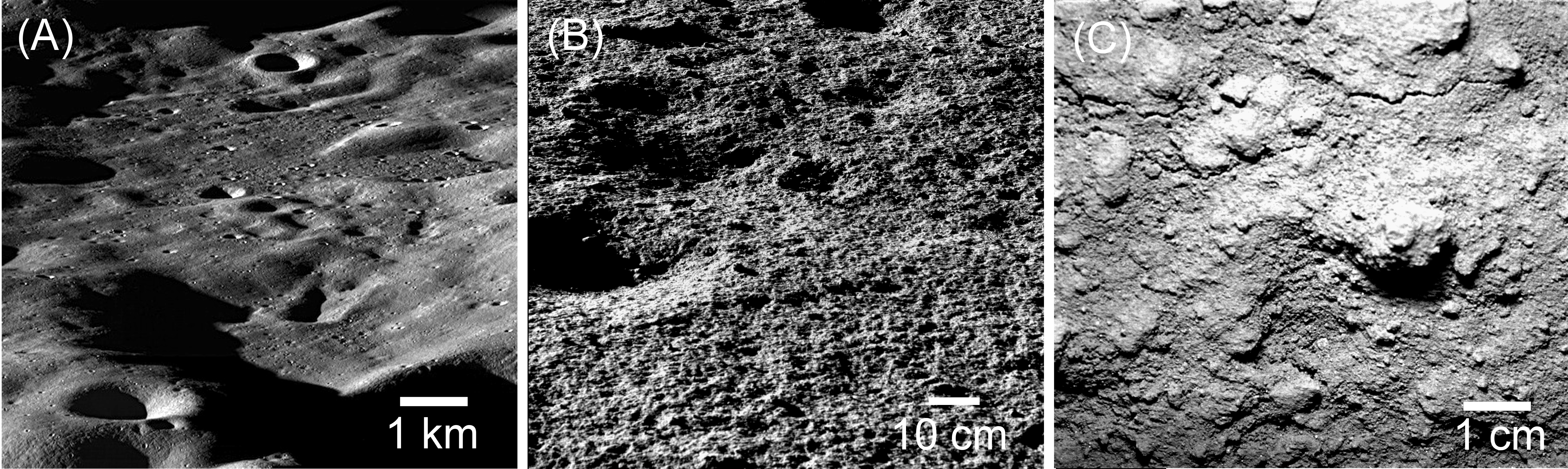}}
\caption{Images reveal shadows on a range of spatial scales: (A) LROC-NAC oblique view over the rim of Cabeus crater near the Moon's south pole. (B) Chang'E-3 close-up surface image taken by the Yutu rover some distance from the landing site. (C) Apollo 14 close-up camera image of undisturbed regolith.}
\label{f:scales}
\end{figure*}

Searches for lunar ice have primarily focused on the large polar craters, where temperatures as low as $\sim$30~K have been measured \citep{paige10a,mazarico11,hayne2015a}. Though lunar thermal models have shown that steep thermal gradients can exist at unresolved spatial scales \citep{buhl68,bandfield15,rubanenko17}, the importance of small-scale shadows for cold-trapping has remained unclear. Here, we report the results of a detailed quantitative study of lunar shadows and cold traps at scales from 1~km down to $<$ 1~cm (Figure~\ref{f:scales}). To do so, we first develop a statistical model of surface topography that is consistent with both the observed instantaneous shadow distribution and the observed temperature distribution. Then, we use theoretical models to connect instantaneous with perennial shadow and temperatures. Small-scale shadows in the polar regions, which we term ``micro cold traps", substantially augment the cold-trapping area of the Moon, and may also influence the transport and sequestration of water.

\section{Shadows}

Here we consider the distribution of shadowed cold traps as a function of their size and latitude. To estimate the fractional surface area $A$ occupied by cold traps with length scales $L$ to $L'$, we calculate the integral
\begin{linenomath*}
\begin{equation}\label{eq:af}
    A(L,L',\varphi) = \int_{L}^{L'}\alpha(l,\varphi)\tau(l,\varphi)dl,
\end{equation}
\end{linenomath*}
where $\alpha(l,\varphi) dl$ is the fractional surface area occupied by permanent shadows having dimension $l$ to $l+dl$, $\tau$ is the fraction of these permanent shadows with maximum temperature $T_{\rm max}<$110~K, and $\varphi$ is the latitude. The problem is then separated into determining $\alpha$ and $\tau$ for each length scale and latitude.

To determine $\alpha(l,\varphi)$, we analyzed high-resolution ($\sim$1~m/pixel) images from the Lunar Reconnaissance Orbiter (LRO) Narrow-Angle Camera (LROC-NAC) \citep{robinson10} and quantified instantaneous shadows. A total of 5250 NAC images acquired with solar incidence angles 70 -- 89\deg\ were analyzed using an automated algorithm to identify shadows and extract their distribution (Methods~B, Figure S4).
The results show that the shadow area fraction increases with incidence angle, and at scales below $\sim100$~m remains approximately constant with scale, while at larger scales it increases (up to the measured maximum scale corresponding to the image size).

To evaluate what portions of instantaneous shadows are permanent (and later, their associated temperatures) a landscape model is needed. We assume a terrain composed of two types of landscapes varying proportions: craters and rough inter-crater plains. The craters are bowl-shaped with variable aspect ratio, and the inter-crater plains are described by a Gaussian surface of normal directional slope distribution parameterized by an root-mean-square (RMS) slope, $\sigma_{s}$ \citep{aharonson06,hayne15b}.

We determined the proportion of crater and inter-crater plains needed to match the measured instantaneous shadow distribution.  For cratered terrain, we derive an analytical relation for the size of shadows in spherical (bowl-shaped) craters, as well as the ratio, $f$, of permanent to instantaneous shadow area (Methods A), and compared with a numerical solution \citep{bussey03}. For the inter-crater terrain, we used ray-tracing to numerically determine instantaneous and permanent shadow statistics for Gaussian surfaces (Methods D). Smith\cite{smith67} developed analytical formulas for the shadow fraction of such surfaces as a function of illumination angle, which compare favorably (within $5\%$) to our numerical model.


The LROC shadow data could not be fit using either the crater or rough surface models alone, but good agreement was obtained with a combination of $\sim$20\% craters by area and $\sim$80\% inter-crater area with $\sigma_{s}$ = 5.7\deg\ (Figure \ref{f:shadow-fraction}). Rosenberg et al.\cite{rosenburg11} found similar slope distributions for the Moon at scales comparable to the NAC images. The Moon's north and south polar regions exhibit an asymmetry in topographic roughness and shadowing, noted previously by Mazarico et al.\cite{mazarico11}. Our permanent shadow distributions agree with those previously mapped on larger scales \citep{mazarico11}, and we also note the topographic dichotomy between the two hemispheres (Figure S6). The south polar region is dominated by several craters $>$10 km in diameter, whereas shadow area in the north is dominated by km-scale and smaller craters. Everywhere, numerous shadows are found down to the smallest resolvable scale, which is $\sim1$~m for LROC-NAC.


\section{Temperatures}

Ice stability is limited by peak surface heating rates, due to the exponential increase in sublimation with temperature. In large shadows, where lateral conduction is negligible, heating is dominated by radiation scattered and emitted by surrounding terrain. We begin by considering this case, and then consider small scales where lateral conduction is important.

We calculated temperatures for both bowl-shaped craters and statistically rough surfaces. Solutions for a bowl-shaped crater were computed using the numerical thermal model of Hayne et al.\cite{hayne2017global}, assuming the analytical irradiance boundary condition of Ingersoll et al.\cite{ingersoll92}. In this case, the temperature within the permanently shadowed portion of the crater depends primarily on the latitude and depth-to-diameter ratio of the crater, $d/D$. Shallower craters have smaller, but colder shadows \cite{ingersoll92}. We considered two log-normal probability functions for $d/D$ with mean $\mu$ and standard deviation $\sigma$. Distribution A  ($\mu$ = 0.14 and $\sigma$ = $1.6\times10^{-3}$) corresponds to a fit to data from Mahanti et al.\cite{mahanti16} using LROC images to derive shapes of craters with $D$ = 10 to 100 m. Distribution B ($\mu$ = 0.076 and $\sigma$ = 2.3$\times10^{-4}$) simulates larger, shallower craters.

\begin{figure}[tbh]
\centering \
\imagetop{\includegraphics[width=0.7\textwidth]{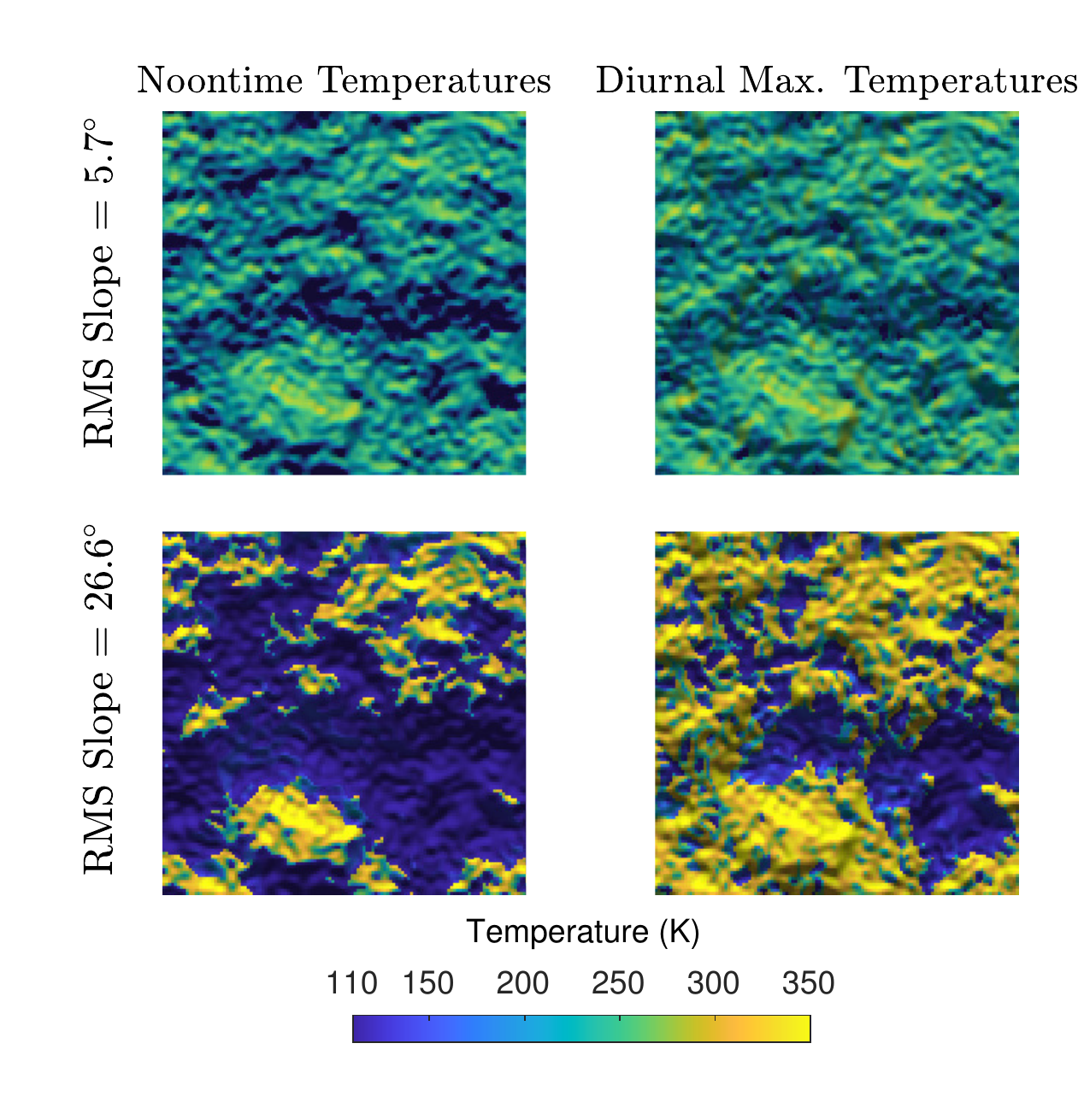}}
\caption{Modeled surface temperatures at 85\deg\ latitude for similar surfaces with two different values of the RMS slope, $\sigma_s$ = 5.7\deg\ (upper panels) and 26.6\deg\ (lower panels). Left-hand panels show peak noontime temperatures, and right-hand panels show the diurnal peak temperatures. In these cases, the model neglects subsurface conduction.}
\label{f:gaussian}
\end{figure}


To estimate shadow fractions and temperatures on rough surfaces, we implemented a numerical model that calculates direct illumination, horizons, infrared emission, visible reflection, and reflected infrared for a three-dimensional topography (Methods D). The RMS slope $\sigma_s$ and solar elevation determine the resultant temperature distribution (Figure ~\ref{f:gaussian}). Rougher surfaces experience more extreme high and low temperatures, but not necessarily larger cold-trapping area; temperatures in shadows may be elevated due to their proximity to steep sunlit terrain. We found the greatest cold-trapping fractional area for $\sigma_s \approx$10~--~20\deg, which is similar to the lunar surface roughness at length scales $\sim$1~cm \citep{helfenstein99}. At the millimeter length scales over which Diviner detects anisothermality \citep{bandfield15}, water ice cold-trapping area is reduced due to surface-to-surface radiative transfer and lateral conduction.

\begin{figure*}[th]
\centering \ \imagetop{\includegraphics[width=0.6\textwidth]{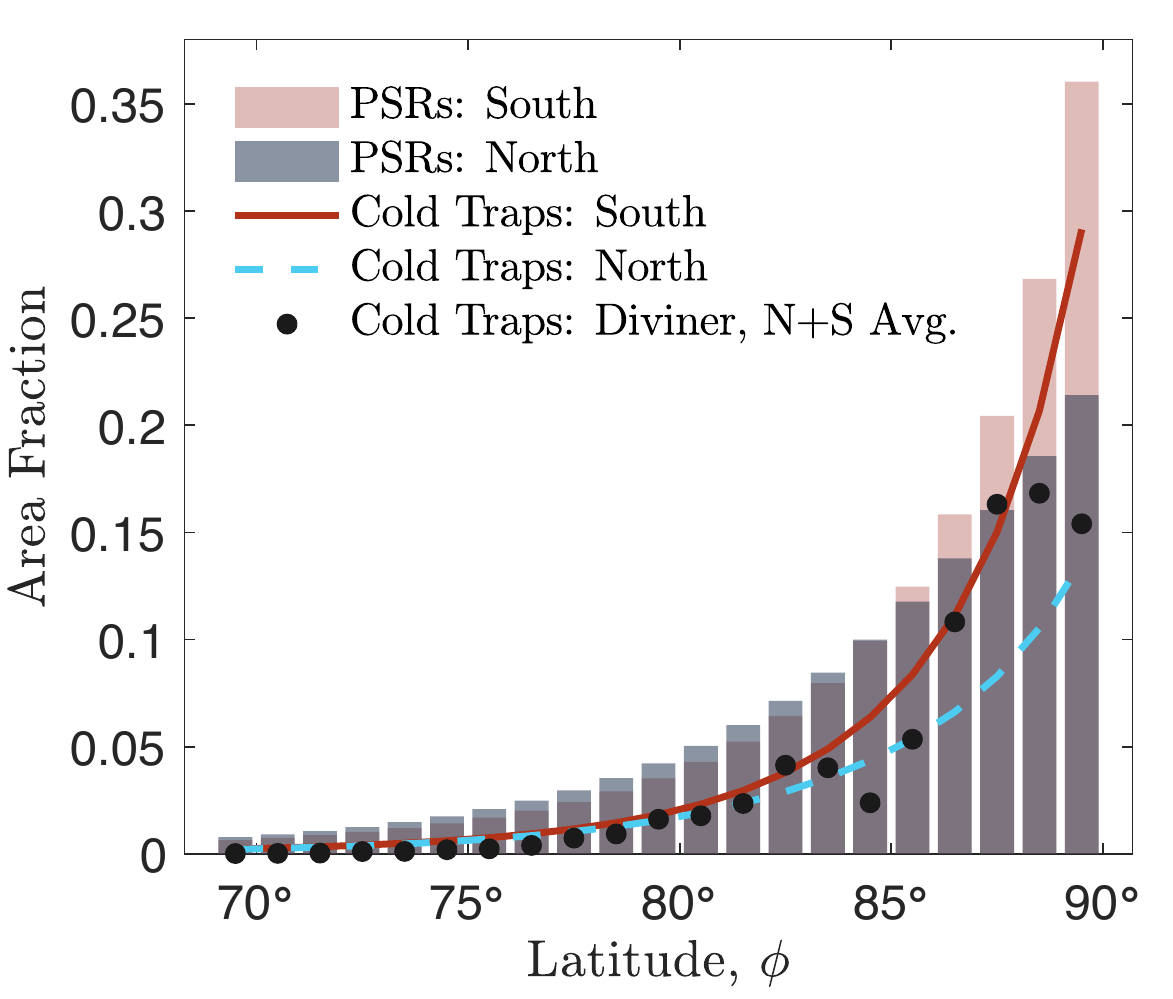}}
\caption{Fraction of total surface area at each latitude remaining perennially below 110 K, the adopted sublimation temperature for water ice. Black points are fractional cold trap areas within 1-degree latitude bands, with temperatures spatially binned at $\sim$250~m. Vertical bars and solid curves are best-fit models of PSR and cold trap area fractions over all spatial scales.}
\label{f:area-fraction}
\end{figure*}

We used thermal infrared emission measurements from the Diviner instrument on board LRO \citep{paige10a} to evaluate peak temperature statistics. Figure~\ref{f:area-fraction} shows that the model reproduces the $\sim$250-m scale Diviner data for crater fractions $\sim$20 -- 50\%, inter-crater RMS slopes $\sim$ 5 -- 10\deg\ and typical $d/D \sim$ 0.08 -- 0.14. These values are consistent with expected surface roughness at similar scales derived from LOLA data \citep{rosenburg11}. Using the highlands median slope of $s_0 = \tan(7.5^\circ)$ with a 17-m baseline, and extrapolating to 250 m using the Hurst exponent $H = 0.95$, we find $s = \tan(7.5^\circ) \left(\text{250~m}/\text{17~m}\right)^{H-1} \approx \tan(6.6^\circ)$. Higher crater densities result in a steeper rise of cold trap area at the highest latitudes, whereas increasing the roughness of the inter-crater plains raises cold trap area more uniformly at all latitudes.

\begin{figure*}[th]
\centering \
\imagetop{\includegraphics[width=0.55\textwidth]{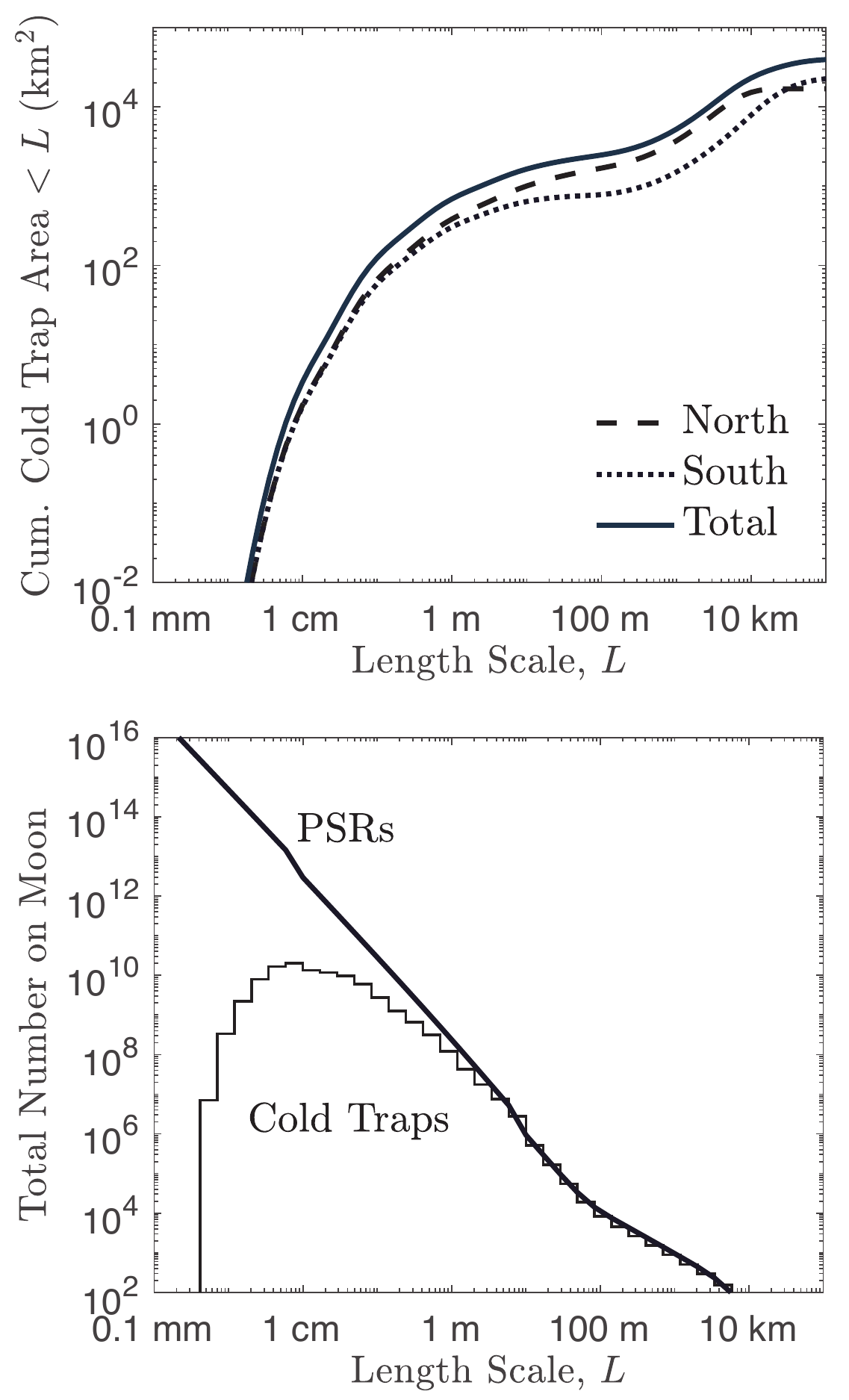}}
\caption{Upper panel: Cumulative area of cold traps ($<$ 110 K) at all latitudes, as a function of shadow length scale, $L$. Lower panel: Modeled number of individual PSRs and cold traps on the Moon. Length scale bins are logarithmically spaced.}

\label{f:cdf}
\end{figure*}

Our model readily allows calculation of both permanently shadowed and cold-trapping areas as a function of size and latitude (Fig. \ref{f:cdf}). Owing to their distinct topographic slope distributions (see above and Fig. S6), the Northern and Southern Hemispheres display different cold trap areas, the south having the greater area overall. This topographic dichotomy also leads to differences in the dominant scales of cold traps: the north polar region has more cold traps of size $\sim$1~m -- 10~km, whereas the south polar region has more cold traps $>$10~km. Since the largest cold traps dominate the surface area, the South has greater overall cold-trapping area ($\sim$23,000 km$^2$) compared to the north ($\sim$17,000 km$^2$). The south-polar estimate is roughly 2$\times$ larger than an earlier $\sim$13,000 km$^2$ estimate derived from Diviner data pole-ward of 85$^\circ$S \citep{paige10a}, due to our inclusion of all length scales and latitudes. About 2,500~km$^2$ of cold-trapping area exists in shadows smaller than 100~m in size, and $\sim$700~km$^2$ of cold-trapping area is contributed by shadows smaller than 1~m in size.

Table~\ref{tbl:watson} and Figure \ref{f:cdf} summarize the PSR and cold trap areas based on the results of this study.
Including seasonal variations, which are neglected here, Williams et al. (2019) \cite{williams19} obtains 13,000~km$^2$ of cold trap area poleward of 80\deg S and 5,300~km$^2$ for the north polar region based on a Diviner threshold of 110~K. Our model shows that many PSRs are not cold traps, particularly those equatorward of 80$^\circ$, which tend to exceed 110 K. Over half a century ago, classical analysis by Watson, Murray and Brown \citep{WMB61a,WMB61b} derived the shadow fraction using photographic data, and assumed a constant $f=0.5$ for the permanent to instantaneous shadow ratio.  We find the overall PSR area fraction is $0.15\%$ of the surface, smaller than the $0.51\%$ found by Watson et al. \cite{WMB61b} (Table~\ref{tbl:watson}). This disagreement is primarily due to the past study assuming a value for $f$ that is substantially higher than what was determined here. As shown in Figure \ref{f:cdf}, we find a large number of PSRs at small scales, extending down to the $\sim 100-\mu$m grain size or smaller.

To determine the minimum size of cold traps, the heat conduction equation including lateral heat transfer is solved (Methods~E). The most numerous cold traps are those of order centimeters, despite being partially warmed by lateral heat conduction (Fig.~\ref{f:cdf}). Continuing down below this length scale, conduction rapidly eliminates cold traps. We note that the more numerous micro cold traps do not dominate in terms of area; for example, those smaller than 1~m account for $\sim$2\% of the total cold trap area, despite being $\sim$100 times more numerous than larger cold traps. The potential volume of the micro cold traps is even smaller, scaling as $\sim D^3$, and we find that those with $D <$1~m could account for $\sim 10^{-5}$ of the total cold-trapping volume, despite being vastly more numerous than larger cold traps. Thus, the potential presence of 10's of meters-thick ice deposits in the Moon's south polar region \citep{rubanenko2019thick} is consistent with our finding that large $>1$~km-scale cold traps are more prevalent in the south than the north, dominating the cold-trapping volume.

\begin{table*}[tb!]
\renewcommand{\arraystretch}{1.1}
\begin{tabular}{|c|c|ccc|c}
\hline
Latitude & PSR  & Noon Shadow  & PSR & Cold trap \\
range (\deg)& area (\%) & area (\%) & area (\%) & area (\%) \\
 & Watson et al.\cite{WMB61b} & & This study & \\
\hline
80--90 & 13.8 & 49 & 8.5 & 6.7 \\
70--80 & 4.3 & 5.5 & 0.5 & 7.0$\times10^{-4}$\\
60--70 & 1.1 & 0.4 & $\sim$0 & $\sim$0 \\
50--60 & 0.5 & $\sim$0 & $\sim$0 & $\sim$0 \\
\hline
Whole Moon & 0.51 & 1.0 & 0.15 & 0.10 \\
\hline
\end{tabular}
\caption{Old and new measurements of PSR and cold trap areas. Percentages are the mean of both hemispheres.}
\label{tbl:watson}
\end{table*}

\section{Conclusions}


More than sixty years after first attempts to quantify the area covered by permanent shadows and ice traps on the Moon \citep{WMB61b}, modern data from LRO and improved models reveal a large number of small PSRs that cumulatively cover a significant area.  We analyzed high-resolution LROC images, and developed models that enable relating instantaneous to permanent shadows. A landscape model consisting of $20 - 50\%$ craters and complementary rough inter-crater plains is simultaneously consistent with three separate measurements: the LROC instantaneous shadow distributions, LOLA terrain roughness properties, and Diviner peak temperatures. We find 0.15\% of the lunar surface is permanently shadowed, with $\sim$10\% of that area distributed in patches smaller than $100~$ m, that is at scales smaller than previously mapped by LOLA topography based illumination models. The most numerous cold traps on the Moon are $\sim$1~cm in scale.


Cold-trapping volatiles in PSRs is limited by the energy input of reflected light and lateral conduction. Thus of the PSR area we find, 0.1\% of the global surface area (roughly 2/3 of PSR area) is sufficiently cold to trap water ice.  Heat diffusion models show that conductive heat becomes significant below decimeter scales on the Moon, and destroys the smallest cold traps $<$1~cm. Nonetheless, the low temperatures of sub-centimeter PSRs may increase the residence time for H$_2$O molecules \cite{poston15,farrell2019young}, influencing their transport and exchange with the lunar exosphere.

The implication of the abundance of small-scale cold traps is that future missions exploring for ice may more easily target and access one of these potential reservoirs. Given the high loss rates due to micrometeorite impact gardening and ultraviolet photodestruction \citep{farrell2019young}, the detection of water within the micro cold traps would imply recent accumulation. If water is found in micro cold traps, the sheer number and topographic accessibility of these locales would facilitate future human and robotic exploration of the Moon.\\


{\bf Corresponding author:}
Paul O. Hayne, Department of Astrophysical \& Planetary Sciences and Laboratory for Atmospheric and Space Physics, 391 UCB, Boulder, CO 80309. \href{mailto:Paul.Hayne@Colorado.edu}{Paul.Hayne@Colorado.edu}\\

{\bf Acknowledgments:}
This study was supported by the Lunar Reconnaissance Orbiter project and NASA's Solar System Exploration Research Virtual Institute.
The authors wish to thank Erwan Mazarico for valuable discussions and data on PSR area derived from LOLA elevation data and illumination models, and Prasun Mahanti for crater depth/diameter ratio data. The authors also wish to thank P. G. Lucey and two anonymous reviewers for insightful criticism that improved this work.
OA wishes to thank the Helen Kimmel Center for Planetary Science, the Minerva Center for Life Under Extreme Planetary Conditions and by the I-CORE Program of the PBC and ISF (Center No. 1829/12).
NS was in part supported by the NASA Solar System Exploration Research Virtual Institute Cooperative Agreement (NNH16ZDA001N) (TREX).\\

{\bf Author contributions:}
PH initiated the study, developed the approach and general methodology (Methods C), analyzed the Diviner data, and performed the model fitting.
OA compiled the shadow fractions from images (Methods B), computed the lateral heat conduction limitation (Method E), and helped construct the overall description of cold trap scale dependence.
NS derived the equations for shadows in a bowl-shaped crater (Methods A) and carried out the numerical energy balance calculations (Methods D).
All authors contributed to the writing of the manuscript.\\

{\bf Data Availability:} All data used in this study are publicly available. The Diviner and LROC data can be accessed through the NASA Planetary Data System: \href{https://pds-geosciences.wustl.edu}{https://pds-geosciences.wustl.edu}. The higher-level data products generated in this study are available from the authors, and will be posted on GitHub: \href{https://github.com/phayne}{https://github.com/phayne}.\\

{\bf Code Availability:} All code generated by this study is available from the authors and/or on GitHub: \href{https://github.com/phayne/heat1d}{https://github.com/phayne/heat1d}
and \url{https://github.com/nschorgh/Planetary-Code-Collection/blob/master/Topo3D}

{\bf Competing interests:} The authors declare no competing financial interests.


\bibliographystyle{naturemag}
\bibliography{planet}

\renewcommand{\deg}{\ensuremath{^\circ}}
\def\deg{\ifmmode{^\circ}\else{$^\circ$}\fi}
\def\h2o{\ifmmode{{\rm H}_2{\rm O}}\else{H$_2$O}\fi}

\section*{Methods}


\setcounter{section}{0}
\renewcommand\thesection{\Alph{section}}

\setcounter{figure}{0}
\renewcommand\thefigure{S\arabic{figure}} 

\section{Relation between permanent and instantaneous shadows in bowl-shaped craters}

Here, an analytical expression is obtained for the size of a shadow in a bowl-shaped (spherical) crater, based on its depth-to-diameter ratio and the elevation of the Sun.
Relations are then derived between the size of the permanent shadow and the noontime shadow.
These results help estimate the size of permanent shadows in craters based on the size of instantaneous shadows in snapshots from orbit.

\begin{figure}[bh!]
\centerline{\includegraphics[width=70mm]{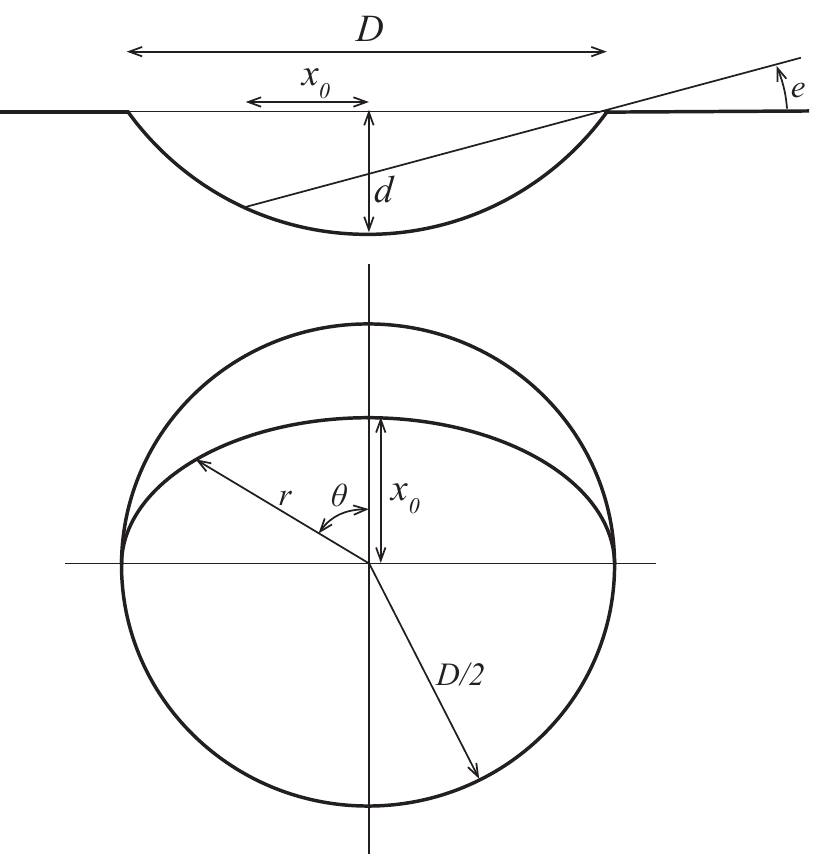}}
\caption{Vertical and horizontal cross sections of spherical crater with shadow, with definitions for variables.}
\label{fig:skizze}
\end{figure}

Figure~\ref{fig:skizze} defines the geometric variables. The crater is a truncated sphere (bowl-shaped), with Diameter $D$, depth $d$, depth to diameter ratio $\gamma = d/D \le 1$, with 
$\beta = 1/(2\gamma) - 2\gamma \geq 0$. The Sun is at elevation angle $e$ and declination $\delta$.

\subsection{Shadow size}

In a Cartesian coordinate system with the $x$-axis in the horizontal plane from the center of the crater towards the Sun, the length of the shadow is obtained after some calculation as
\begin{equation}
\frac{D}{2} + x_0 = D \cos e \left[ \cos e - \frac{\beta}{2} \sin e \right]
\end{equation}
and in terms of the unitless coordinate $x_0'=2x_0/D$,
\begin{equation}
x'_0 = \cos^2 e - \sin^2 e -\beta \cos e \sin e.
\end{equation}

A transect along the direction of a Sun ray that does not pass through the center of the crater is geometrically similar, and hence
\begin{equation}
x=x'_0 \sqrt{\left(\frac{D}{2}\right)^2 - y^2} 
\end{equation}
so the shadow boundary is part of an ellipse.
Normalized by the crater area $A_{\rm crater}=\pi D^2/4$, the area of the shadow is
\begin{eqnarray}
A_{\rm shadow} \over A_{\rm crater} &=& {1+x'_0 \over 2 }  =  ( \cos e - \frac{\beta}{2} \sin e ) \cos e 
\end{eqnarray}
The illuminated area is the complement of this shadowed area.
$A_{\rm shadow}>0$ implies $\tan e < 2/\beta$.
If in equilibrium with sunlight, the temperature in the shadow is known analytically\cite{buhl68,ingersoll92}.

\subsection{Smallest shadow throughout solar day}

\subsubsection{Simple case: The pole}

At the pole, the permanent shadow is circular,
\begin{eqnarray}
A_{\rm permanent} \over A_{\rm crater} &=& x'^2_0,
\label{eq:Apolar} \\
{A_{\rm permanent} \over A_{\rm noon} } &=& { 2x'^2_0 \over x'_0+1 }
\label{eq:ApolaroverAnoon}
\end{eqnarray}
where $A_{\rm noon}$ is measured when the Sun is highest (at solstice).
%
%
For small declination, (\ref{eq:Apolar}) and (\ref{eq:ApolaroverAnoon}) become
\begin{eqnarray}
A_{\rm permanent} \over A_{\rm crater} &\approx &  1-2\beta\delta 
\label{eq:perm_noon} \\
{A_{\rm permanent} \over A_{\rm noon} } &\approx& 1 - \frac{3}{2} \beta \delta
\label{main1}
\end{eqnarray}

There can be instantaneous shadow without permanent shadow. Permanent shadow requires $x'_0>0$,
\[
\beta < \frac{1}{\tan e} - \tan e
\]
\[
\tan e < -\beta/2 + \sqrt{\beta^2/4+1}
\]
For comparison, the criterion to have any shadow is $\beta < 2/\tan e$.

\subsubsection{Elevation of Sun over time}
At latitude $\varphi$, the elevation of the Sun is related to its azimuth $a_s$ by
\begin{equation}
\cos a_s = {\sin\delta - \sin\varphi \sin e \over \cos\varphi \cos e}
\end{equation}
$a_s = \pi-\vartheta_s$, such that $\cos a_s = -\cos\vartheta_s$.
For small $\delta$, small $e$, and $\varphi$ close to the pole ($\sin\varphi  \approx 1$, $\cos\varphi  \approx \pi/2-\varphi$),
\begin{equation}
e \approx \delta +  (\pi/2-\varphi) \cos\vartheta_s
\label{eapprox}
\end{equation}

\subsubsection{Shadow length in polar coordinates}

In polar coordinates,
$x=r\cos\vartheta$, $y=r\sin\vartheta$,
\begin{equation}
r=\frac{D}{2} {x'_0 \over \sqrt{\cos^2\vartheta + x'^2_0\sin^2\vartheta}}
\label{rpolar}
\end{equation}
The direction of the Sun can be incorporated by a shift in $\vartheta$,
\begin{equation}
r=\frac{D}{2} {x'_0(\vartheta_s) \over \sqrt{\cos^2(\vartheta+\vartheta_s) + x'^2_0(\vartheta_s)\sin^2(\vartheta+\vartheta_s)}}
\end{equation}
where $x'_0$ is a function of $\vartheta_s$ because the elevation of the Sun depends on $\vartheta_s$. To find the shortest shadow, which does not occur at the same time along  different directions,
$d r/d\vartheta_s =0$. This leads to
\begin{equation}
\frac{dx'_0}{de} \frac{de}{d\vartheta_s} = x'_0(x'^2_0-1) \tan(\vartheta+\vartheta_s) 
\label{mincond}
\end{equation}

\subsubsection{Some perturbative results}

\begin{figure}[tbh!]
\centering \
\includegraphics[width=70mm]{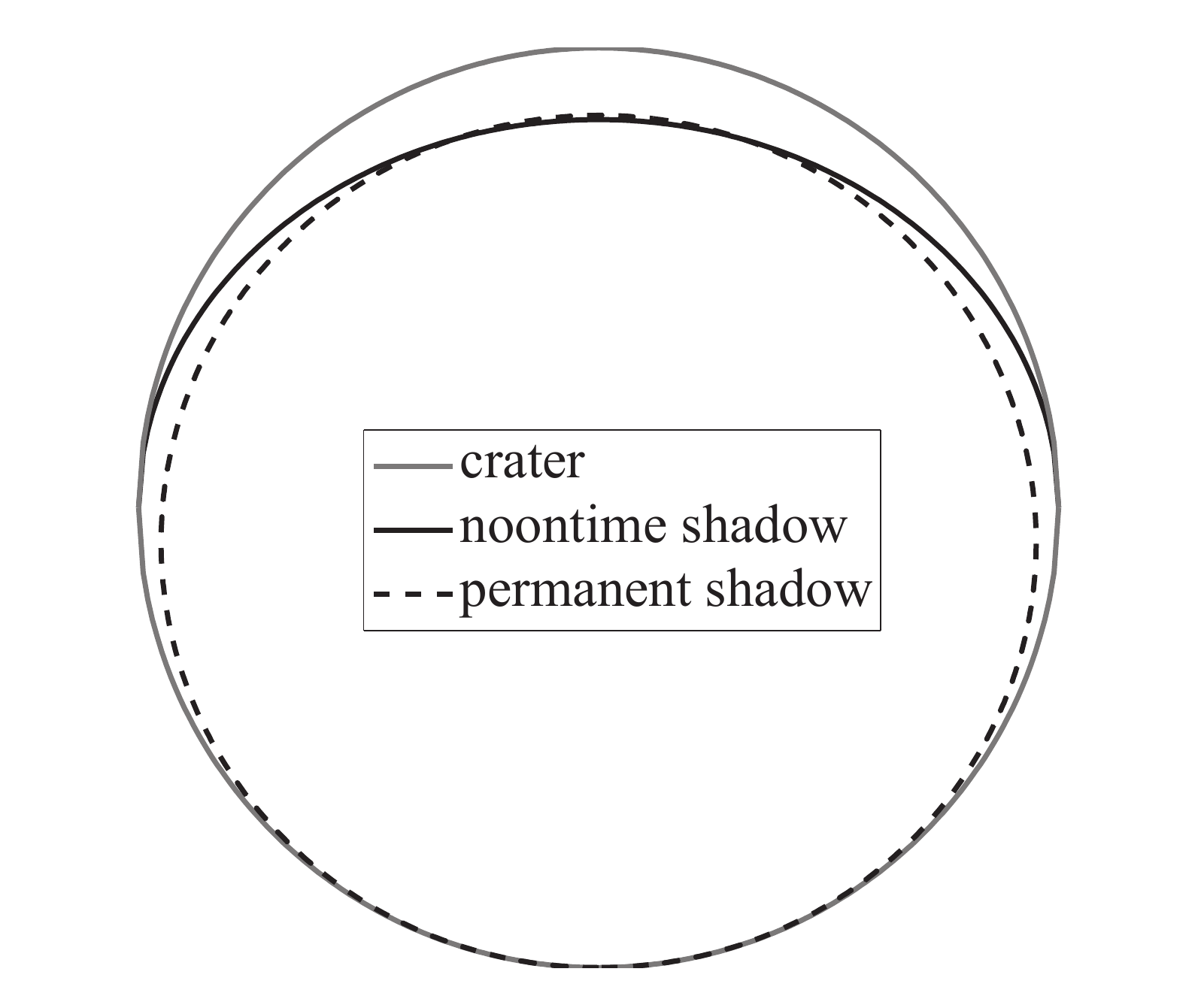}
\caption{Top view of circular crater with the exact noontime shadow boundary (\ref{rpolar}) and the approximate extent of permanent shadow (\ref{rmin}). The diameter to depth ratio of the crater is 5 and the maximum elevation of the Sun, $e_0$, is 4$^\circ$.}
\label{fig:shadow}
\end{figure}

The approximate size of permanent shadow is obtained for small $e$,
\begin{equation}
x'_0 \approx 1 - \beta e  - 2e^2
\end{equation}
\begin{equation}
\frac{dx'_0}{de} {1\over x'_0(x'^2_0-1) } \approx {1\over 2e } + B
\end{equation} 
with $B= \frac{1}{\beta} + \frac{3\beta}{4} $.
From (\ref{eapprox}),
\[
\frac{de}{d\vartheta_s} = - \left(\frac{\pi}{2}-\varphi \right)  \sin \vartheta_s 
\]
The condition for the minimum (\ref{mincond}) becomes
\begin{equation}
- e_0 \sin \vartheta_s
\left( {1\over 2 e_0  \cos \vartheta_s + 2\delta} + B \right)
= \tan(\vartheta+\vartheta_s) 
\label{min_approx}
\end{equation}
where $e_0 =  \pi/2-\varphi$ is the co-latitude.
For $\delta=0$ and small $B$ (\ref{min_approx}) becomes
\begin{equation}
{\tan\vartheta_s \over 2}  = - \tan(\vartheta+\vartheta_s) 
\label{eq:eq}
\end{equation}
leading to the solution
\begin{equation}
\tan\vartheta_s = \frac{3}{2} \cot\vartheta - \sqrt{2 + \frac{9}{4} \cot^2\vartheta}
\end{equation}
This is approximated with $\vartheta_s \approx - \vartheta/2$, because it satisfies (\ref{eq:eq}) for $\vartheta=0$ as well as for $\vartheta\to\pi$.
(However, for small $\vartheta$ the solution is $\vartheta_s\approx -\frac{2}{3}\vartheta$.)
The minimum shadow length along each direction $\vartheta
$ is approximately,
\begin{equation}
r_{\rm min} \approx \frac{D}{2} {x'_0(-\vartheta/2) \over \sqrt{\cos^2(\vartheta/2) + x'^2_0(-\vartheta/2)\sin^2(\vartheta/2)}}
\label{eq:rmin_full}
\end{equation}
Within this approximation, $x'_0 \approx 1- \beta  e$ and $e \approx e_0 \cos\vartheta_s$,  where $e_0$ is now both co-latitude and the highest elevation of the Sun.
Equation (\ref{eq:rmin_full}) becomes
\begin{equation}
r_{\rm min} \approx \frac{D}{2} \left[ 1 - \beta e_0 \cos^3 \left(\frac{\vartheta}{2} \right) \right]
\label{rmin}
\end{equation}
Figure~\ref{fig:shadow} shows this result.

The area of permanent shadow is 
\begin{eqnarray}
A_{\rm permanent} &\approx& 
\frac{1}{2} \int_{-\pi}^\pi r^2_{\rm min} d\vartheta 
\approx \left(1 - \frac{8 \beta e_0}{3\pi}\right) A_{\rm crater} 
\label{eq:Aperm}
\end{eqnarray}
This equation also provides an estimate for the condition of permanent shadow: $\beta e_0 < 3\pi/8$. To the same order of approximation,
\begin{eqnarray}
A_{\rm instantaneous} \over A_{\rm crater} &=& \frac{1}{2}(x'_0+1)  \approx \left( 1-\frac{\beta e}{2} \right)
\label{eq:Ainst}
\\
{A_{\rm permanent} \over A_{\rm instantaneous} } &\approx& 1 - \beta \left(\frac{8}{3\pi}e_0 - \frac{1}{2}e \right)
\\
{A_{\rm permanent} \over A_{\rm noon} } &\approx& 1 - \beta e \left(\frac{8}{3\pi} - \frac{1}{2} \right)\approx 1 - 0.35 \beta e
\label{main2}
\end{eqnarray}
This result is for zero solar declination and for small Sun elevation $e$ (high latitude).

\begin{figure}[tb!]
\centering \
\includegraphics[width=70mm]{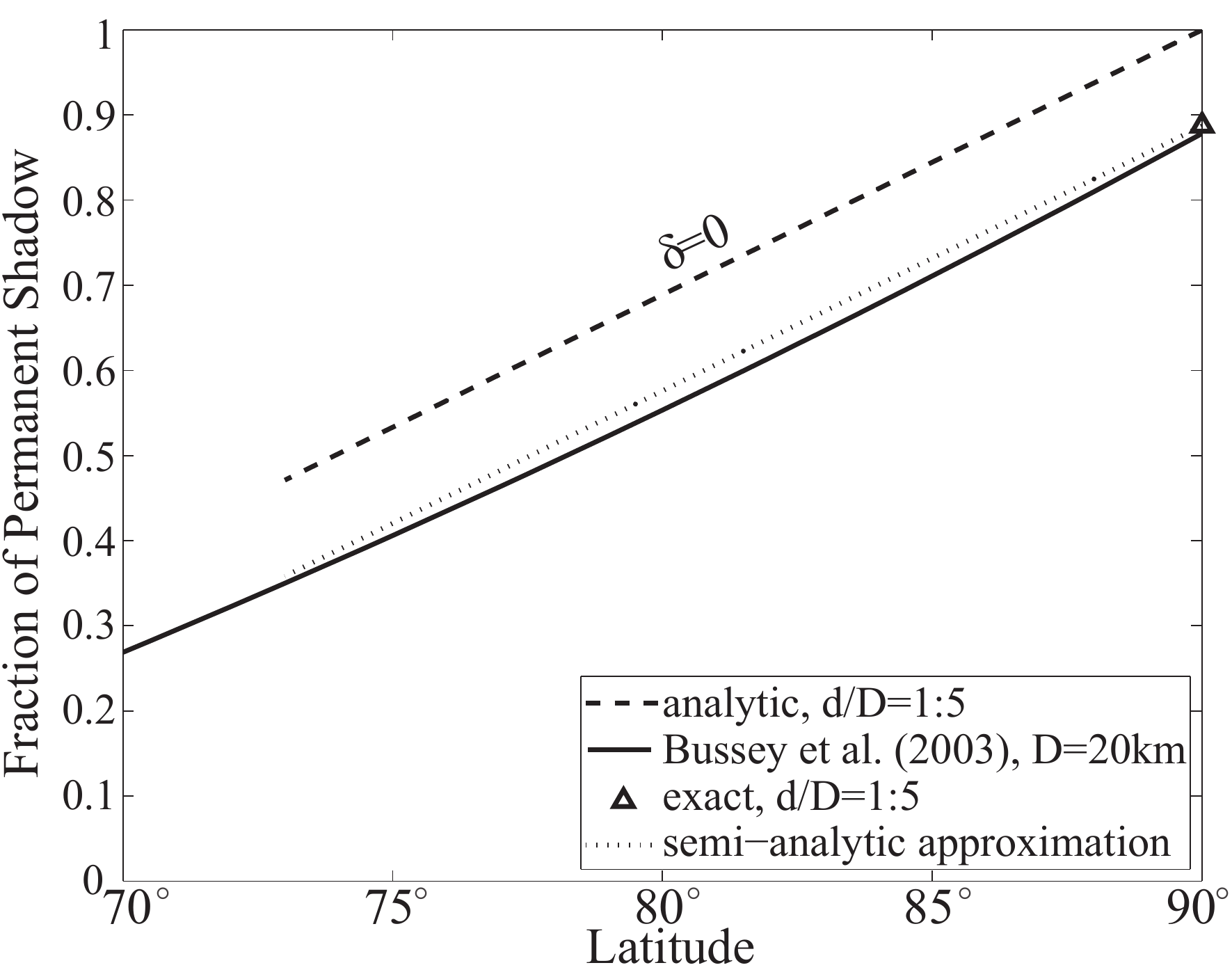}
\caption{Comparison between the numerical results of Bussey et al.\cite{bussey03}, the analytic result for zero declination (\ref{eq:Aperm}), and approximation (\ref{eq:semi}). 
The triangle shows the analytic result for the pole (\ref{eq:Apolar}).}
\label{fig:bussey}
\end{figure}

Figure~\ref{fig:bussey} shows a comparison between the analytic expression and numerical results by Bussey et al.\cite{bussey03} that are based on the crater shapes by Pike\cite{pike77}.  The perturbative expansion accurately captures the latitude dependence, and the offset is due to the declination effect.

\subsubsection{Declination effect}

The comparison in Fig.~\ref{fig:bussey} suggests that the declination effect can be taken into account by subtracting (\ref{eq:perm_noon}) from (\ref{eq:Aperm}).
Empirically,
\begin{equation}
{A_{\rm permanent} \over A_{\rm crater}} \approx 
1 - \frac{8 \beta e_0}{3\pi} - 2 \beta\delta .
\label{eq:semi}
\end{equation}
This agrees well with the numerical results (Fig.\ \ref{fig:bussey}).
The declination effect is larger than would have been estimated by merely adding it to the maximum Sun elevation.
For the noontime or instantaneous shadow, however, exactly this can be done,
and (\ref{eq:Ainst}) remains valid.

$\delta$ in (\ref{eq:semi}) should be chosen as the maximum declination.
For the practical purpose of estimating permanent shadow size from instantaneous shadow size,
\begin{equation}
f_c \equiv {A_{\rm permanent} \over A_{\rm instantaneous} }\approx 
1 - \beta \left( \frac{8}{3\pi} e_0  + 2 \delta_{\mathrm max} - \frac{1}{2} e \right)
\label{eq:final}
\end{equation}
where $e_0$ is the co-latitude and $e$ is the instantaneous Sun elevation ($\pi/2$ minus the incidence angle).
Practically, $A_{\rm instantaneous}$ can be determined as a function of incidence angle, and then (\ref{eq:final}) is used to estimate the size of permanent shadows in craters as a function of latitude.

\section{Shadow Measurement}

We used publicly available image data from the Lunar Reconnaissance Orbiter Camera (LROC) Narrow Angle Camera (NAC) to estimate instantaneous shadow areas, over a range of solar incidence angles.
Our algorithm identifies contiguous regions of similar brightness in each grayscale image with known pixel scale. Shadows are easily distinguishable from illuminated regions, due to the high dynamic range of the NAC images and natural contrast of the Moon.
We surveyed 5250 images distributed such that there are hundreds of images in each latitude/incidence angle computation bin.  Each image's pixel brightness distribution was fit by the sum of two gaussian functions.  The peak centered on the darker pixel values corresponds to the shadow areas, and the shadow threshold was extracted as three gaussian half-widths above the mean of this peak. Visual inspection of multiple images was used to verify shadows are correctly identified by this algorithm \citep{aharonson17}. We then extracted spatially connected components in the binary shadow image using a standard flood-fill algorithm for detection of pixels with shared edges, and compiled the area distribution of these components.  The area $A_i(\theta)$ of each individual shadowed region $i$ in an image with solar incidence angle $\theta$ and area $A_{\rm image}$ is calculated based on the pixel scale and number of contiguous pixels contained in the region. The linear dimension of a shadowed region is $L_i = (A_i/\pi)^{1/2}$, and the fractional shadow area from $L$ to $L+\delta l$ is \begin{equation}
M(L,\theta) \delta L= \frac{1}{A_{\rm image}}
\Sigma_i A_i(L < L_i < L+\delta L).
\end{equation}


\section{Scale Dependence of Shadow and Cold Trap Areas}

To calculate the fractional area occupied by cold traps, it is necessary to determine the functions $\alpha(l,\varphi)$ and $\tau(l,\varphi)$ from equation (\ref{eq:af}). Here, $\alpha(l,\varphi) dl$ is the fractional surface area occupied by permanent shadows having dimension $l$ to $l+dl$, $\tau$ is the fraction of these permanent shadows with maximum temperature $T_{\rm max} <$ 110~K, and $\varphi$ is the latitude.

\begin{figure}[bh!]
\centering \
\includegraphics[width=70mm]{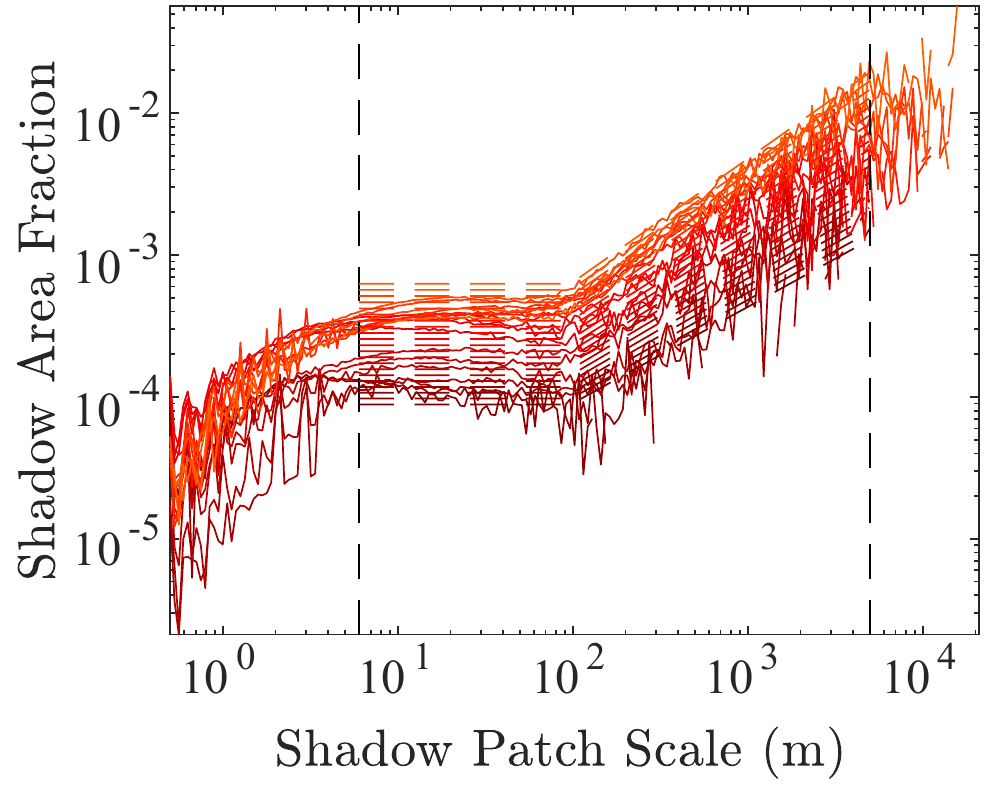}
\caption{Instantaneous shadow fraction from LROC-NAC images for a range of solar incidence angles (68\deg, 69\deg, ..., 88\deg, lighter tones representing increasing incidence angle), binned in logarithmically spaced shadow patch scales from 0.5 m to 50 km. Dashed lines represent a fit to the data.}
\label{fig:ishadow}
\end{figure}

\begin{figure*} [th]
\centering \ 
\includegraphics[width=70mm]{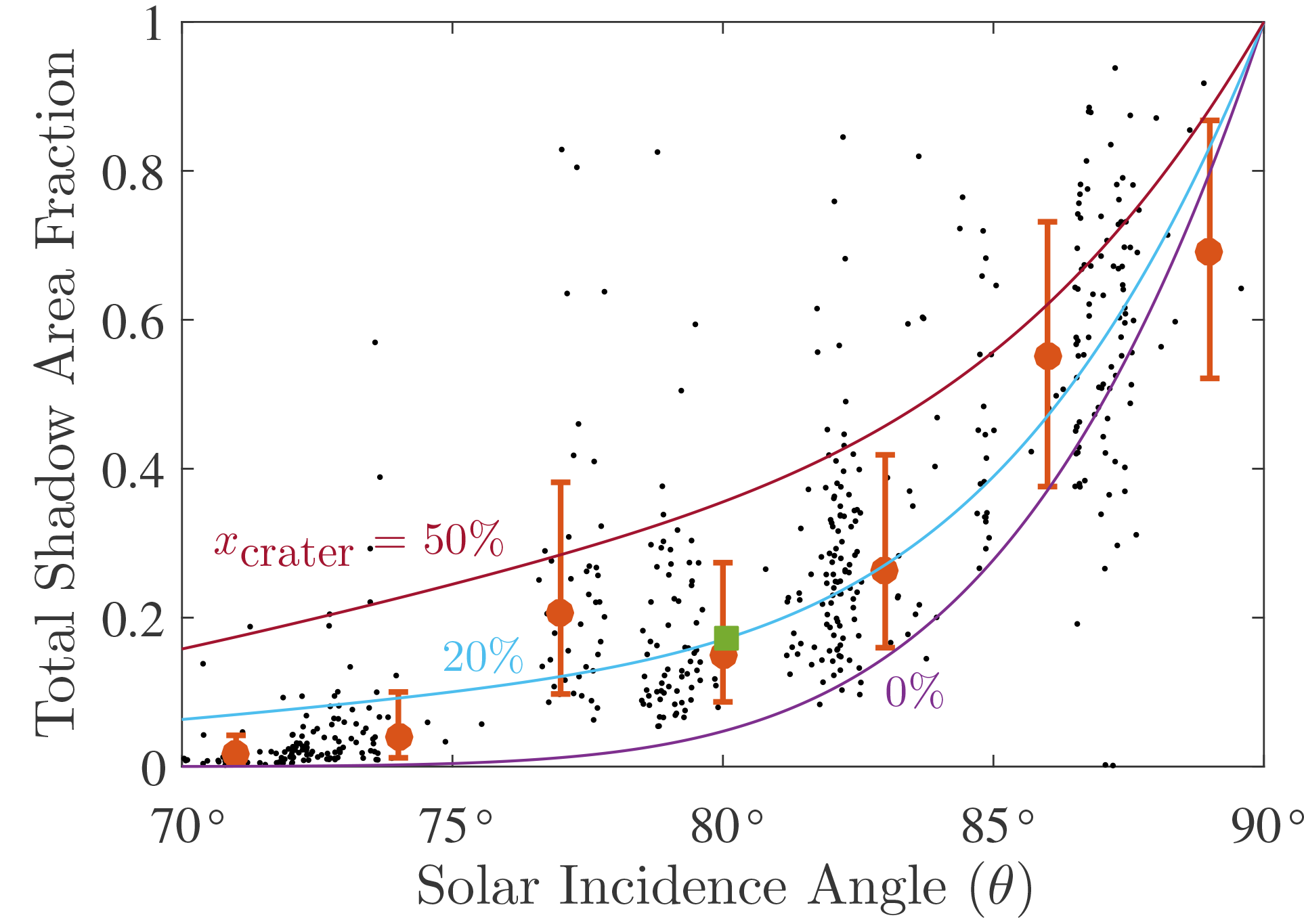}
\caption{Comparison of shadow fraction from LROC image data (points) to the model (curves). Error bars indicate the mean and standard deviations within each solar incidence angle bin, and $x_\mathrm{crater}$ is the area occupied by craters within the rough terrain.}
\label{f:shadow-fraction}
\end{figure*}

\begin{figure}[bh!]
\centering \
\includegraphics[width=70mm]{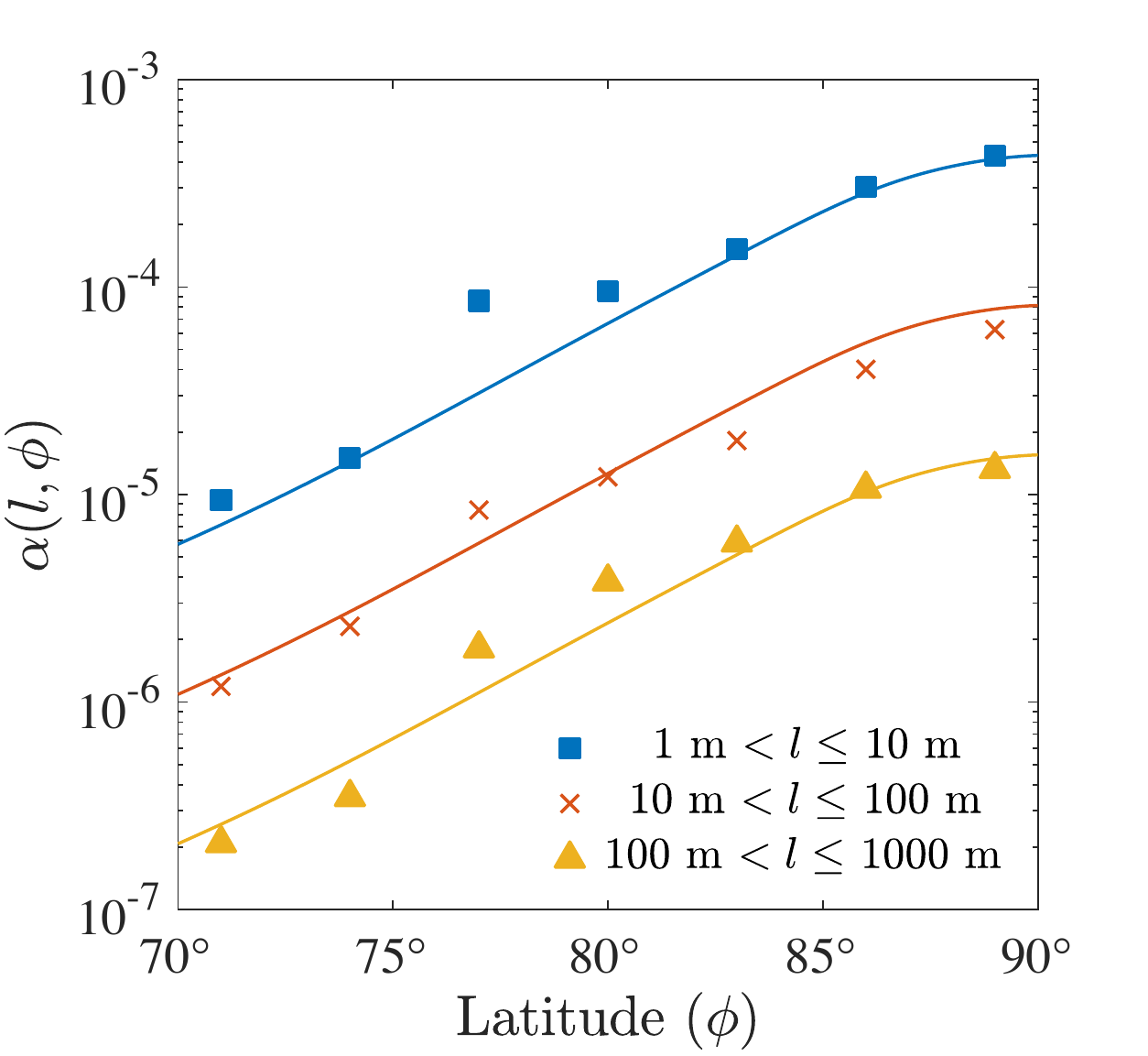}
\caption{Fraction $\alpha$ (units of m$^{-1}$) of the total surface area occupied by permanent shadows as a function of latitude $\varphi$ and length scale $l$. The model curves shown assume a crater fraction of 20\% and inter-crater plains with RMS slope $\sigma_\mathrm{s} = 5.7^\circ$}
\label{fig:alpha}
\end{figure}

First, to determine $\alpha$, we tabulated values of the fractional area of instantaneous shadow from LROC images, as described above. This shadow  fraction is then used in the integral

\begin{equation}
    M(L,\varphi) = \int_{L}^{L+\delta L} \frac{\alpha(l,\varphi)}{f(l, \varphi)} dl
\end{equation}
where $\delta L \ll L$, and $f(l,\varphi)$ is the ratio of permanent to instantaneous noontime shadow, which also may depend on length scale and latitude. Figure \ref{fig:ishadow} shows the instantaneous shadow fraction over a range of scales and incidence angles, along with our model fit: 
\begin{equation}
    \log_{10}M_\textrm{fit}(L,\varphi) = B_0 + B_1\cos\varphi + B_2\log_{10}L + B_3\cos\varphi \log_{10}L
\end{equation}
with best-fit coefficients $B_0=-4.89, B_1=-1.38, B_2=0.89, B_3=-0.57$. The model also fits the observed break in slope (Fig. \ref{fig:ishadow}) at $L_\textrm{break}$ by forcing $M(L,\varphi) = M(L_\textrm{break},\varphi)$ for all $L \le L_\textrm{break}$. From the LROC-NAC data used in this study, we find $L_\textrm{break} \approx 100$~m.

We assume that $\alpha$ is not very sensitive to small changes in $l$, i.e., we observe that shadow areas increase in proportion to logarithmic size bins. In this case, over a restricted range of $l$ from $L$ to $L + \delta L$,
\begin{equation}
    \alpha(L,\varphi) = \frac{\partial}{\partial L}(fM) \approx f(L,\varphi)\frac{M(L+\delta L,\varphi)-M(L,\varphi)}{\delta L}
    \label{eq:alpha}
\end{equation}
Figure \ref{fig:alpha} displays the resulting best-fit values of $\alpha$ for a range of solar incidence angles and three length scales. These were derived using (\ref{eq:alpha}) and the shadow data from LROC, with $f$ calculated using (\ref{eq:final}) and crater depth/diameter ratio $\gamma = 0.1$.



\begin{figure}[tbh!]
\centering \
\includegraphics[width=70mm]{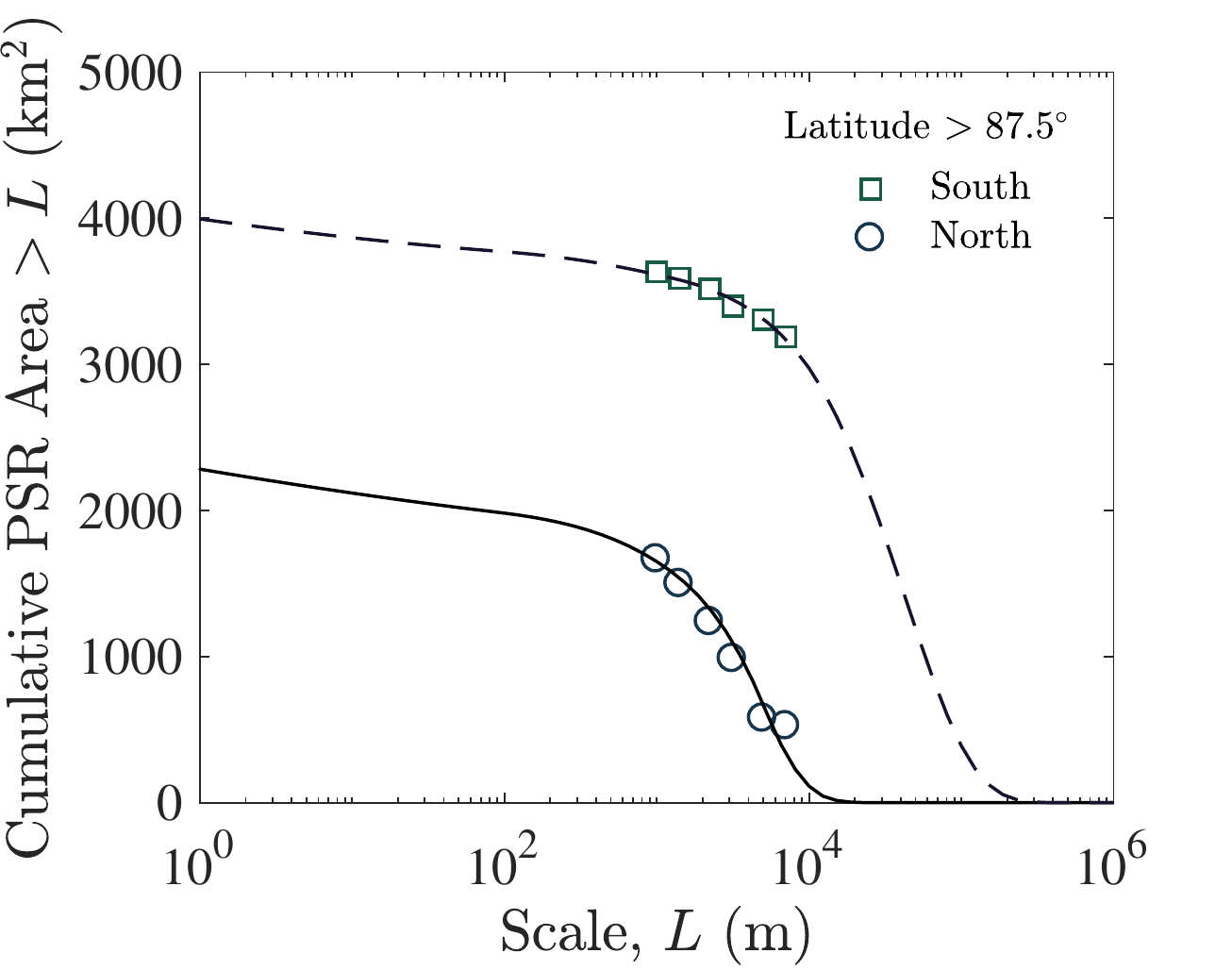}
\caption{Cumulative fraction of the total surface area occupied by permanent shadows with length scale $l > L$. Data points are from Mazarico et al.\cite{mazarico11}, for each polar region, $>87.5$\deg.}
\label{fig:mazarico_fits}
\end{figure}

The next step is to determine the fraction $\tau$ of these permanent shadows that are cold traps. We considered two regimes: (1) shadows large enough that conduction from warm sunlit surfaces is negligible; (2) shadows small enough to be affected by lateral heat conduction. In the first case, surface temperatures in shadows are determined by the incident radiation, which we calculate exactly. For the case of small shadows, the $\tau$ term is indeed affected by lateral heat conduction, which we estimate as follows.

\subsection{Craters}

\begin{figure}[tbh!]
\centering \
\includegraphics[width=70mm]{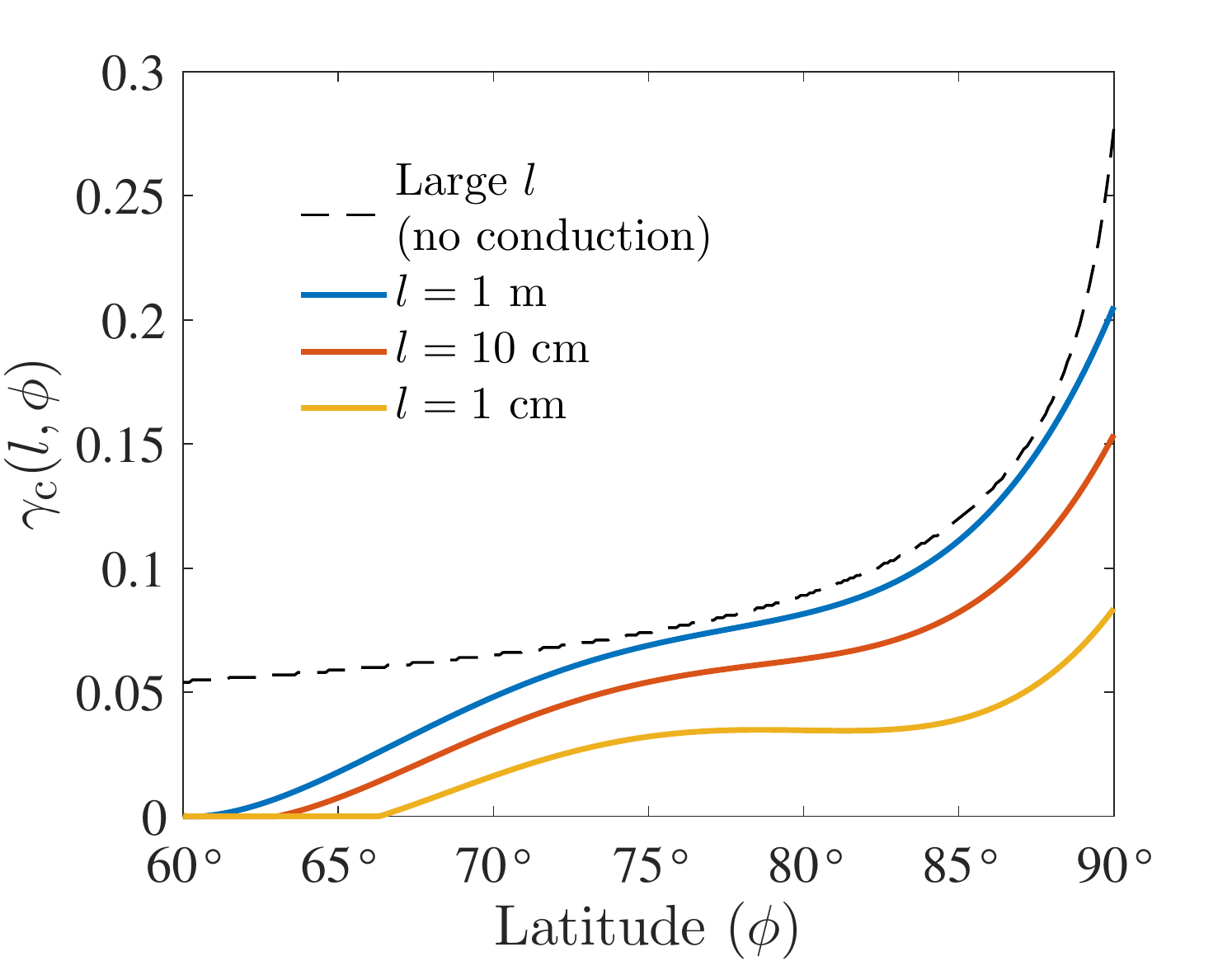}
\caption{Critical depth/diameter ratio $\gamma_\textrm{c}$ of craters, for which $\gamma < \gamma_\textrm{c}$ indicates the PSR is a cold trap for water ice. }
\label{fig:gamma}
\end{figure}

For each latitude $\varphi$ and length scale $l$, there exists a maximum depth/diameter ratio $\gamma_\textrm{c}(l,\varphi)$ corresponding to craters whose permanently shadowed portions have $T_\textrm{max} < 110$~K. Since shallower craters have colder PSRs \citep{ingersoll92,vasavada99}, the criterion $\gamma < \gamma_\textrm{c}$ is sufficient to determine whether a PSR is a cold trap. For large $l$, $\gamma_\textrm{c} \rightarrow \gamma_\textrm{c,0}$, the value absent conduction. In determining $\gamma_\textrm{c}$ for smaller $l$, we used a 2-d heat conduction model (described below) to estimate the contribution of lateral conduction into PSRs. Figure \ref{fig:gamma} shows a summary of these results. 

The 2-d model indicates that conduction eliminates cold traps with sizes ranging from $\sim1$~cm near the pole, to $\sim10$~m at 60\deg\ latitude. We note that this size range depends on the choice of $T_\textrm{max}$ for cold-trapping, and neglects multiple-shadowing.

$\gamma_\textrm{c,0}(\varphi)$ is calculated from the analytical boundary conditions derived by \cite{ingersoll92}, coupled to the 1-d thermal model. The results of this 1-d transient model are generally consistent with the 2-d steady-state model.  Using the modeled values of $\gamma_\textrm{c}$, we then calculate
\begin{equation}
    \tau_\textrm{c}(l,\varphi) = \int_{0}^{\gamma_\textrm{c}(l,\varphi)} P(\gamma)d\gamma
\end{equation}
where $P(\gamma)$ is the log-normal probability distribution function for crater depth/diameter ratio $\gamma$: 
\begin{align}
    P(\gamma; \mu, \sigma) &= \frac{1}{\gamma s \sqrt{2\pi}} e^{-\frac{(\ln \gamma - m)}{2s^2}} \\
    m &= \ln\left( \frac{\mu}{\sqrt{1+\frac{\sigma}{\mu^2}}}\right) \\
    s^2 &= \ln\left( 1 + \frac{\sigma}{\mu^2}\right)
\end{align}

\begin{figure}[tbh!]
\centering \
\includegraphics[width=70mm]{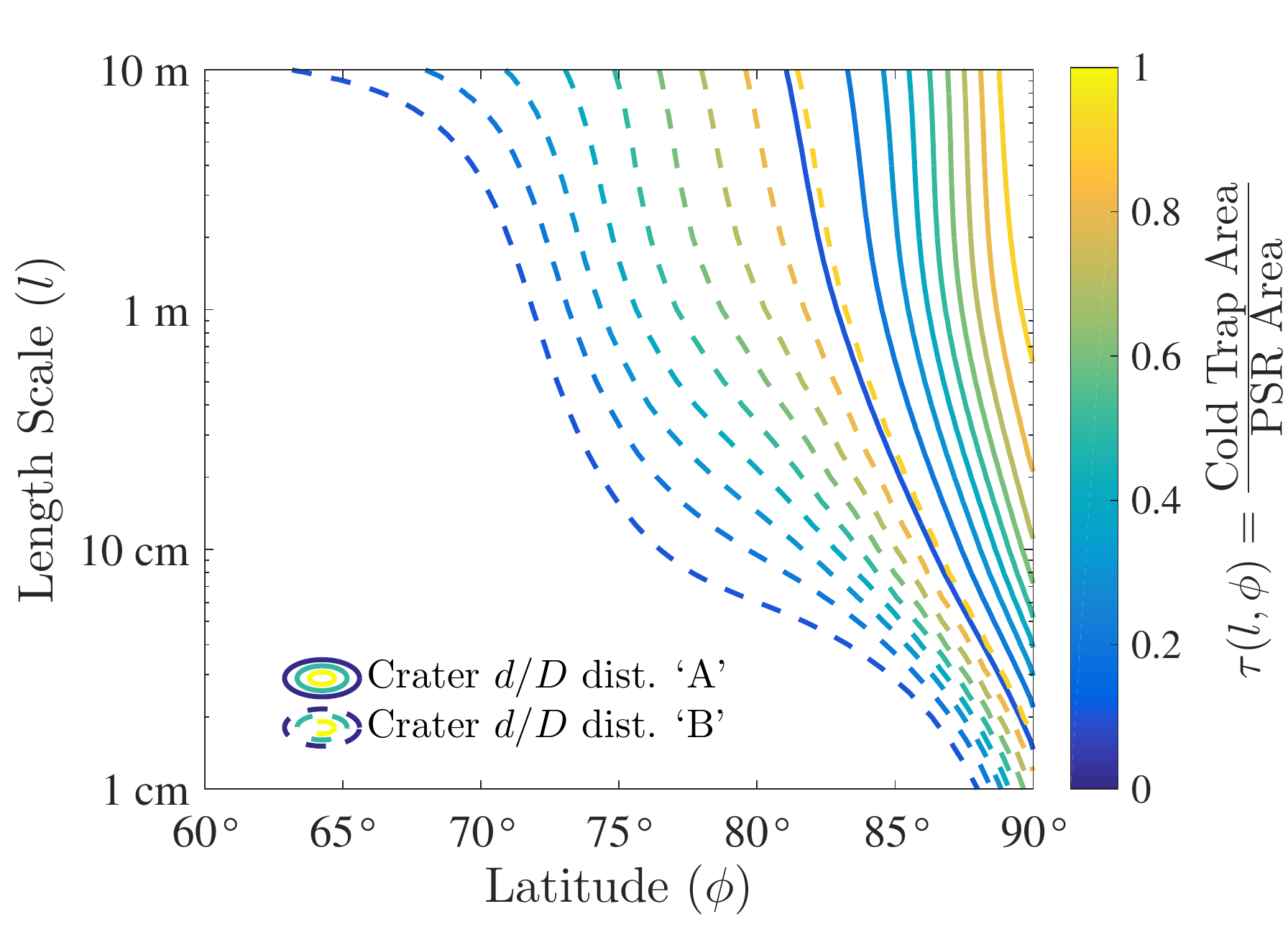}
\caption{Fraction of permanently shadowed regions (PSR) inside craters that are cold traps for water ice, $T_{\rm max} < 110$~K. Results are shown for two log-normal probability distributions `A' (deeper craters, $\mu = 0.14$) and `B' (shallower craters, $\mu = 0.076$). Contours are plotted for $\tau = 0.1, 0.2,...,0.9$. }
\label{fig:tau}
\end{figure}

Figure \ref{fig:tau} displays results of this calculation for the two log-normal probability distributions `A' (deeper craters, $\mu = 0.14$) and `B' (shallower craters, $\mu = 0.076$).

\subsection{Rough Surface}
Although we did not explicitly model lateral conduction for the Gaussian rough surface, our model results absent conduction provide an upper limit on the relative cold-trapping area, which we call $\tau_\textrm{p,0}(\sigma_\textrm{s};\varphi)$. This quantity (the ratio of cold-trap area to PSR area, absent conduction) is scale-independent, but instead depends on the RMS slope, $\sigma_\textrm{s}$. We modeled the latitude- and scale-dependence of lateral conduction assuming it is the same for rough surfaces as for craters.

\subsection{Distribution Functions}

A number of useful measures of cold trap and PSR area can be determined once $\alpha$ and $\tau$ are determined. Defining the fractional cold trap area per unit length $l$,
\begin{equation}
    A_l \equiv \frac{\partial A(\varphi,l)}{\partial l} = \alpha(\varphi,l)\tau(\varphi,l)
\end{equation}
and
\begin{equation}
    A_{l\varphi} \equiv \frac{\partial^2 A(\varphi,l)}{dl d\varphi} = \alpha(\varphi,l)\tau(\varphi,l)\cos\varphi.
\end{equation}
Integrating these differential density functions can provide the areas of cold traps, such that for example, the cumulative distribution function in a hemisphere is:
\begin{equation}
    \textrm{CDF}_{<x} = \int_0^{\pi/2} \int_0^x A_{l\varphi}dld\varphi.
\end{equation}
The length scale $l$ may be thought of as the effective radius of the shadow patch, such that for example, for circular areas of diameter $D$, $l^2=D^2/4$. The number density is related to the area density by 
\begin{equation}
    N_l =  \frac{A_l}{\pi l^2}  .
\end{equation}
The number density of PSRs is similarly
\begin{equation}
    N_\textrm{PSR},l = \frac{\alpha(\varphi,l)}{\pi l^{2}}.
\end{equation}
Given a hemispherical volume $V(l) = (2/3)\pi l^3$, the total volume per unit area of cold traps in a hemisphere with dimensions from $L$ to $L'$ is
\begin{equation}
    V(L,L') = \frac{2}{3}\int_0^{\pi/2} \int_L^{L'} A_{l\varphi}l dl d\varphi
    = \frac{2\pi}{3}\int_0^{\pi/2} \int_L^{L'} N_{l} l^3 dl d\varphi.
\end{equation}


\section{Energy Balance on Rough Surface}

A numerical model is used to calculate direct and indirect solar irradiance on arbitrary topography.
The model code and model documentation are available online \citep{github17}.

Gaussian surfaces have been created for RMS slope values of 0.1 (5.7\deg), 0.3 (16.7\deg), and 0.5 (26.7\deg) and a Hurst exponent of 0.9, according to the following procedure \citep{lahi57b}:
\begin{enumerate}
    \item Assign random phases to each element in Fourier space, observing the symmetry that the Fourier transform of a real function must have.
    \item The Fourier amplitudes are assigned according to a power law with the desired exponent.
    \item Beyond a wave number threshold (for short wavelengths) the amplitudes are set to zero. The Fourier amplitudes of the longest wavelengths are also set to zero, so the resulting topography will result in more than just a single hill or valley. 
    \item The field is inverse Fourier transformed into real space, resulting in a surface with a Gaussian distribution for elevation and derivatives.
    \item Derivatives and RMS slope are then calculated in real space, and all heights are multiplied by a factor to achieve the desired RMS slope.
\end{enumerate}
Horizons are determined by using rays, every 1\deg\ in azimuth, and the highest horizon in each direction is stored. The direct solar flux at each surface element defines instantaneous and permanent shadows. 
The field of view for each surface element is calculated in terms of the spherical angle as viewed from the other element, and stored.  Mutual visibility is determined by calculating the slope of the line that connects the two elements and comparing it to the maximum topographic slope along a ray in the same direction, tracing outward (ray casting).

With this geometric information the direct and scattered fluxes can be calculated as a function of time (sun position). The scattering is assumed to be Lambertian, and the Sun a point source. The incoming flux determines the equilibrium surface temperature, which in turn is used to evaluate the infrared fluxes in the same way as the scattered short-wavelength flux. An albedo of 0.12 and an emissivity of 0.95 are assumed.
Numerical results for the temperature field compare favorably with the analytical solution for a bowl-shaped crater \citep{ingersoll92}.

Equilibrium surface temperatures are calculated over one sol (lunation) for various solar elevations, including scattering of visible light and infrared emission between surface elements, calculated as described above. 
Shadows and surface temperatures were calculated for Gaussian surfaces at latitudes of 70--90\deg\ and solar declinations of 0\deg\ and 1.5\deg.
The spatial domain consists of 128$\times$128 pixels, as much larger domains would have required excessive computation time. When evaluating the results, a margin is stripped from each of the four sides of the domain to eliminate boundary effects.

\section{Lateral Heat Conduction}

An analytical solution is available for a disk of diameter $D$ at temperature $T_1$ surrounded by an infinite area at temperature $T_2$ in cylindrical geometry. This heat flux is\citep{janna99} $F=2Dk(T_2-T_1)$ , where $k$ is the thermal conductivity. Likewise the flux into a hemisphere at fixed temperature buried in a semi-infinite medium is $F=\pi D k(T_2-T_1)$, where $D$ is now the diameter of this sphere.
However these solutions involve a significant flux at the temperature discontinuity.

\begin{figure}[bht!]
\centering \
\begin{tabular}{ll}
    \imagetop{\bf a)}   &
\imagetop{\includegraphics[width=70mm]{figures/puck1eT.eps}}  \\
    \imagetop{\bf b)} & \imagetop{\includegraphics[width=70mm]{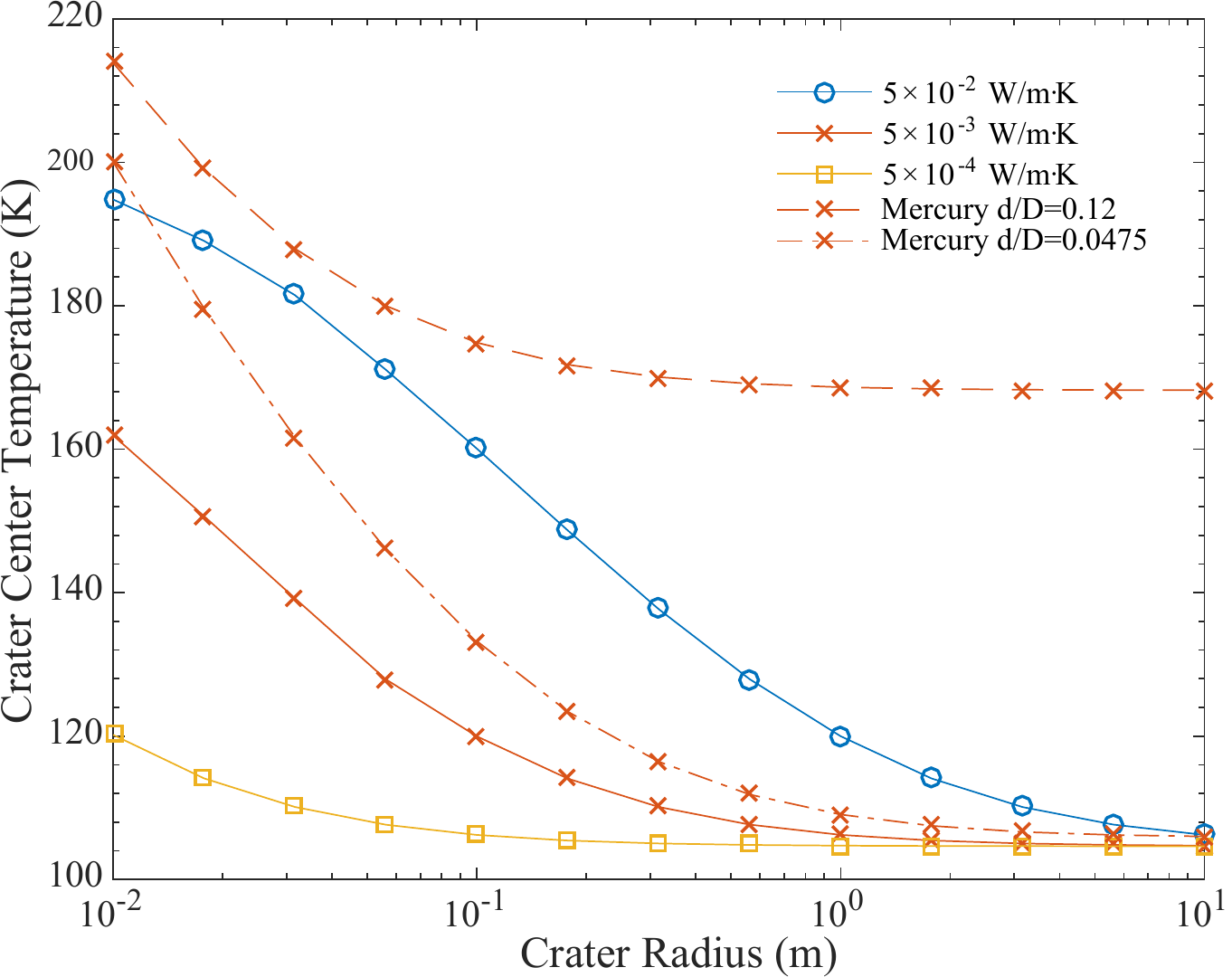}}
    \end{tabular}
\caption{Solution to the heat equation in a cylindrically symmetric geometry. a) Cross section through the domain showing temperature contours and heat flow directions. b) Temperature at the center of a crater (representing the minimum temperature) for different thermal and orbital assumptions.\label{f:model2d}
}
\end{figure}

A better estimate is obtained by numerically solving the cylindrically symmetric Laplace equation with radiation boundary conditions at the surface and no-flux boundary conditions at the lateral and bottom boundaries. This static solution uses the mean diurnal insolation as boundary condition, which is an accurate approximation for length scales $>7$~cm, comparable to the diurnal thermal skin depth \citep{hayne2017global}. To determine the effects of temperature oscillations on shadows smaller than the skin depth, we used the 1-d model described in Methods C. Within a disk of unit radius, the equilibrium flux calculated with the analytic solution for a bowl-shaped is used as incident flux. The domain needs to be chosen large enough to accurately represent the heat flux from the surrounding into the shadowed region (Figure~\ref{f:model2d}).

\end{document}